\documentclass[USenglish,twocolumn]{article}
\usepackage[usenames]{color}

\usepackage[utf8]{inputenc}
\usepackage[big]{dgruyter}

\begin{document}

  \articletype{Research Article{\hfill}Open Access}

  \author*[1,2]{Kargaltseva N.~S.}

\author[1,2]{Khaibrakhmanov S.~A.}
\author[1,2]{\fbox{Dudorov A.~E.}}
\author[1]{Zamozdra S.~N.}
\author[3]{Zhilkin A.~G.}


  \title{\huge Influence of the magnetic field on the formation of protostellar disks}

  \runningtitle{Article title}


  \begin{abstract}
{
We numerically model the collapse of magnetic rotating protostellar clouds with mass of 10~$M_{\odot}$. The simulations are carried out with the help of 2D MHD code Enlil. The  structure of the cloud at the isothermal stage of the collapse is investigated for the cases of weak, moderate, and strong initial magnetic field. Simulations reveal the universal hierarchical structure of collapsing protostellar clouds, consisting of flattened envelope with the qausi-magnetostatc disk inside and the first core in its center. The size of the primary disk increases with the initial magnetic energy of the cloud. The magnetic braking efficiently transports the angular momentum from the primary disk into the envelope in the case, when initial magnetic energy of the cloud is more than 20\,\% of its gravitational energy. Intensity of the outflows launched from the region near the boundary of the first core increases with initial magnetic energy. The `dead' zone with small ionization fraction, $x<10^{-11}$, forms inside the first hydrostatic core and at the base of the outflow. Ohmic dissipation and ambipolar diffusion determine conditions for further formation of the protostellar disk in this region.
}
\end{abstract}
  \keywords{magnetic fields, magnetic-gas-dynamics (MHD), numerical simulation, star formation, interstellar medium}

  \journalname{Open Astronomy}
\DOI{DOI}
  \startpage{1}
  \received{..}
  \revised{..}
  \accepted{..}

  \journalyear{2014}
  \journalvolume{1}

\maketitle


{ \let\thempfn\relax
\footnotetext{\hspace{-1ex}{\Authfont\small \textbf{Corresponding Author: Kargaltseva N.~S.:}} {\Affilfont Chelyabinsk state university, 129 Br. Kashirinykh str., Chelyabinsk 454001, Russia; Ural Federal University, 51 Lenina str., Ekaterinburg 620000, Russia; Email: kargaltsevans@mail.ru}}
}

{ \let\thempfn\relax
\footnotetext{\hspace{-1ex}{\Authfont\small \textbf{Khaibrakhmanov S.~A.:}} {\Affilfont Chelyabinsk state university, 129 Br. Kashirinykh str., Chelyabinsk 454001, Russia; Ural Federal University, 51 Lenina str., Ekaterinburg 620000, Russia}}
}

{ \let\thempfn\relax
\footnotetext{\hspace{-1ex}{\Authfont\small \textbf{\fbox{Dudorov A.~E.}:}} {\Affilfont Chelyabinsk state university, 129 Br. Kashirinykh str., Chelyabinsk 454001, Russia; Ural Federal University, 51 Lenina str., Ekaterinburg 620000, Russia}}
}

{ \let\thempfn\relax
\footnotetext{\hspace{-1ex}{\Authfont\small \textbf{Zamozdra S.~N.:}} {\Affilfont Chelyabinsk state university, 129 Br. Kashirinykh str., Chelyabinsk 454001, Russia}}
}

{ \let\thempfn\relax
\footnotetext{\hspace{-1ex}{\Authfont\small \textbf{Zhilkin A.~G.:}} {\Affilfont Institute of Astronomy of the Russian Academy of Sciences (INASAN), Moscow, 119017, Russia}}
}

\section{Introduction}

Observations show that stars form in cold dense cores of filamentary interstellar molecular clouds. This protostellar clouds (PSCs hereafter) have typical density $n=10^3-10^6$~cm$^{-3}$, temperature $T=10-20$~K, size $R=0.03-0.3$~pc, mass $M=0.1-30 \,M_{\rm \odot}$, velocity dispersion $\sigma \le 3$~km~s$^{-1}$, and angular velocity $\omega =10^{-12}-10^{-14}$~s$^{-1}$~(see, e.g., \cite{Carey1998, BerginTafalla2007, Morii2021} and also review by~\citet{DudKhaibr2017}). 

The distribution of the specific angular momentum inside PSCs was studied only on large scales of $0.1$~pc. The energy of rotation of PSCs is of the order of several percent of the gravitational energy~\citep{Goodman1993, Caselli2002}.

Measurements of the Zeeman effect in the OH lines indicates that PSCs have magnetic field with strength of $B=10^{-5}-10^{-4}$~G. According to polarization  mapping of PSCs, the magnetic field has hourglass morphology~\citep{Girart2006, Crutcher2012, Li2021}.

It is believed that gravitationally unstable PSCs undergo gravitational collapse and evolve into class 0 young stellar objects (YSOs hereafter) observed as the IR sources and interpreted as protostars surrounded by dense flattened envelope of gas and dust~\citep{Andre1993}. Bipolar outflows are the characteristic feature of class 0 YSOs~\citep{Myers1988, Andre1995, Galametz2020}.

Studies in the submillimeter range show that the envelopes of class 0 YSOs have size of $500 - 12000$~au. The envelopes are flattened along the axis of rotation and/or the magnetic field direction~\citep{Wiseman2001, Maureira2020}. High angular resolution studies of the central regions around protostars in recent years revealed small disks with signs of Keplerian rotation and sizes of $2 - 300$~au~\citep{Ohashi2014, Dunham2014, Persson2016, Pineda2019, Tobin2020}. These disks can be associated with  protostellar accretion disks. 

The geometry of the magnetic field is quasi-radial and partially quasi-toroidal inside the envelopes of class~0 YSOs~\citep{Lee2019, HullZhang2019}. 
The angular momentum distribution changes across disk to envelope transition at radii from 1000 to 10000~au~\citep{Goodman1993, Ohashi1997, Caselli2002, Belloche2013}. 

The transition state between the PSC and class 0 YSO is still not possible to observe. In order to determine the conditions for the formation of protostellar disks, it is important to perform numerical modelling of the initial stages of the collapse of PSCs and investigate the formation and evolution of the internal structure of the PSC.

In the simulations of the collapse of magnetic rotating PSCs in the ideal MHD approximation, disks are not formed or very small geometrically thick disks are formed, which does not agree with the observations of protostellar disks~\citep{MellonLi2008}. This is due to the fact that the frozen-in magnetic field transports the specific angular momentum too quickly from the central part of the cloud. This so-called catastrophic magnetic braking can be weakened due to the action of magnetic ambipolar diffusion, Ohmic dissipation, and/or turbulence allowing for the formation of the protostellar disks~\citep{BlackScott1982, Mouschovias1991, HennebelleCiardi2009, Tsukamoto2017, Zhao2020}. However, the exact conditions for the formation of protostellar disks are not determined because of large difference between the initial conditions and numerical models of the PSC. 
Modern simulations are mainly concentrated on the accretion stages of the collapse of solar mass PSC~\citep[see, e.g.][]{Hennebelle_Fromang2008, Zhao2020}. 
Investigation of the initial stages of collapse will make it possible to more accurately determine the conditions for the formation of protostellar disks and come closer to solution of the problem of magnetic braking catastrophe. 

Previously, we numerically simulated the isothermal collapse of rotating magnetic PSCs and found that the hierarchical structure of the cloud forms at this stage~\citep{Khaibrakhmanov2021, Kargaltseva2021}. In this paper, we further develop our approach and analyse the efficiency of the magnetic braking at the isothermal stage of the collapse by simulating the collapse for various initial magnetic energies of the cloud. In Section~2, we describe the problem setup and the code Enlil used for simulations. General picture of the hierarchical structure of the collapsing PSC is outlined in Section~\ref{sec:hierarchy}. Dynamics of the collapse in clouds with different initial magnetic energies is investigated in Section~\ref{sec:mf}. Section~\ref{sec:diff} analyzes  the role of the Ohmic dissipation and magnetic ambipolar diffusion in the evolution of the magnetic flux of the cloud. Section~4 contains main conclusions and discussion of the results.

\section{Problem statement and numerical code}

We consider a homogeneous spherically symmetric rotating magnetic PSC with mass $M_0 = 10\,M_{\odot}$ and radius $R_0 = 0.1$~pc. The rotation axis coincides with the initial magnetic field direction. We adopt axial symmetry approximation and use the cylindrical coordinates, $(r,\, 0, z)$, where $r$ is the radial distance from the cloud's center, $z$ is the height above its equatorial plane.

Initial state of the cloud is determined by the ratios of thermal, rotational, and magnetic energies of the cloud to the modulus of its gravitational energy: $\varepsilon_{\rm t}$, $\varepsilon_{\rm w}$, and $\varepsilon_{\rm m}$, respectively.

The simulations are carried out using the two-dimensional MHD code Enlil~\citep{Dud1999, Dud1999b}. To solve the equations of ideal MHD, the code utilizes the quasi-monotone TVD scheme of the third order of approximation in the spatial variable and the first order of approximation in time. Poison's equation is solved with an implicit alternating direction method. The generalized Lagrange multiplier method is used to clean up the divergence of the magnetic field. An adaptive moving mesh is used with $150 \times 150$ cells. 
Ohmic dissipation and ambipolar diffusion are taken into account in the model~\citep[see][]{zhilkin_et_al_2009}, and the ionization fraction is calculated following~\cite{DudSaz1987}. 
Thermal evolution of the cloud is modelled using a variable effective adiabatic index, $\gamma_{\rm eff}$, in the same manner as in~\cite{Kargaltseva2021}. In the process of isothermal collapse, $\gamma_{\rm eff}=1.001$ is used in the equation of state of the ideal gas. When the density reaches $\rho_{\rm c}= 10^{-13}$ g~cm$^{-3}$, the effective adiabatic index switches to the value of $5/3$. This corresponds to the moment of the formation of the first hydrostatic core~\citep{larson1969}. Similar approach was used, for example, by~\citet[][]{MasunInuts2000}. 

\section{Results}

In order to investigate the influence of the magnetic field on the collapse, we set the thermal and rotational parameters to $\varepsilon_{\rm t}=0.3$ and $\varepsilon_{\rm w}=0.01$, and varied the magnetic parameter in simulations. Three simulation runs were performed with  $\varepsilon_{\rm m}=0.01$ (run I), $0.2$ (run II) and $0.6$ (run III).  The first case corresponds to highly non-equilibrium cloud, $\varepsilon_{\rm t} + \varepsilon_{\rm w} + \varepsilon_{\rm m}\ll 1$ with weak magnetic field. The cloud with moderate magnetic field strength is considered in the second run. The third run studies the nearly equilibrium cloud, $\varepsilon_{\rm t} + \varepsilon_{\rm w} + \varepsilon_{\rm m}\lesssim 1$, with strong magnetic field. Adopted values of the magnetic parameter correspond to the following values of non-dimensional mass-to-flux ratio: $\lambda=10.5$ (Run I), $2.4$ (Run II) and $1.4$ (Run III).

In the following, we use the characteristic time of the collapse of the magnetic rotating PSC $t_{\rm fmw}=t_{\rm ff}\left(1 - \varepsilon_{\rm m} - \varepsilon_{\rm w}\right)^{-1/2}$ as a unit of time, where $t_{\rm ff}$ is the free fall time~\citep[see][]{Khaibrakhmanov2021}.

\subsection{Hierarchical structure of the collapsing protostellar cloud}
\label{sec:hierarchy}

First of all, let us discuss general picture of the isothermal collapse of the magnetic rotating PSC using run II as a reference. This run with moderate magnetic field has been analysed in detail in our previous paper~\citep{Kargaltseva2021}.

\begin{figure}
   \centering
   \includegraphics[width=0.49\textwidth]{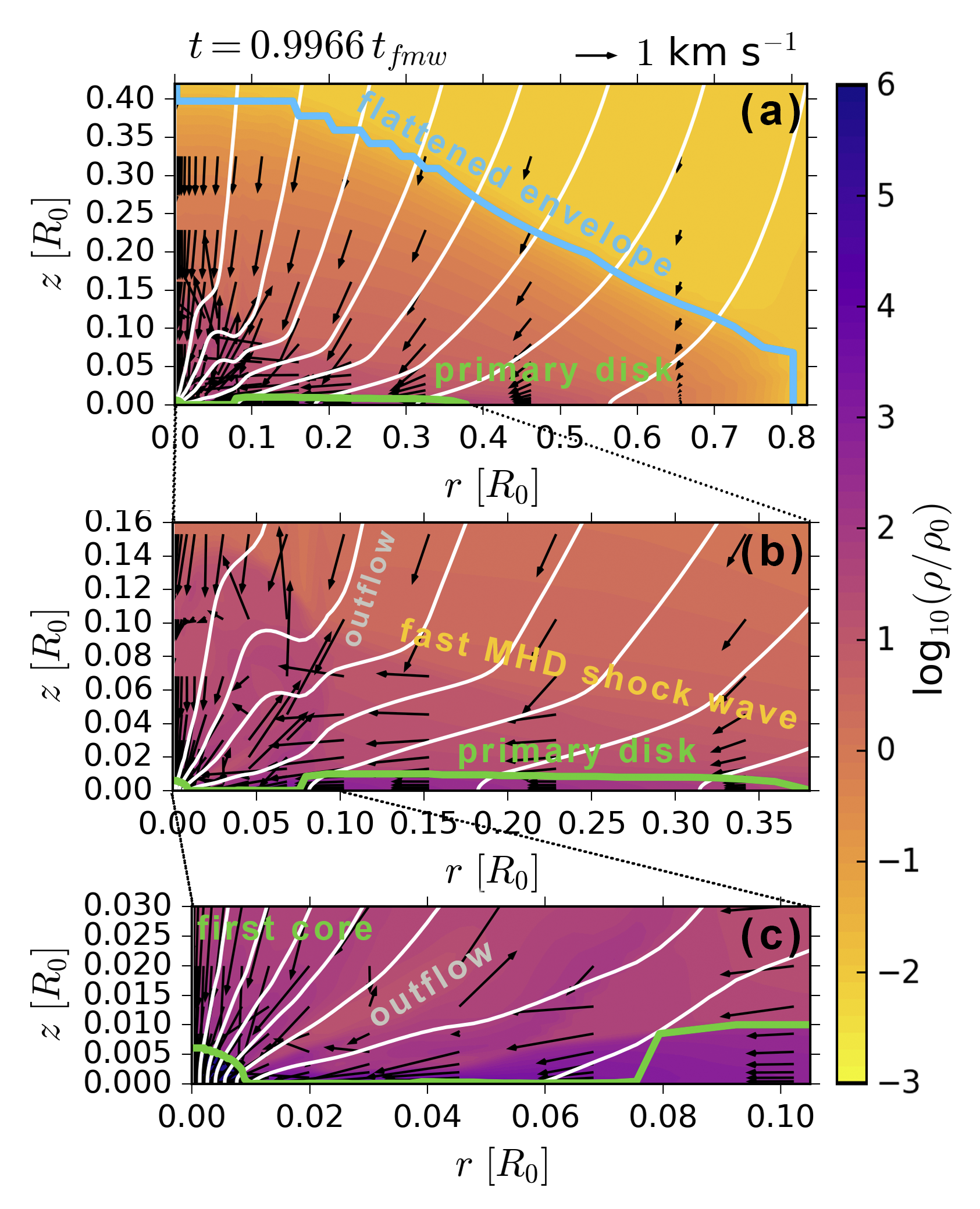}
   \caption{Distribution of density (color filling), velocity field (arrows) and poloidal magnetic field (white lines) in the simulation of the collapse of the magnetic rotating PSC of mass $10\,M_{\odot}$ with initial radius $R_0 = 0.1$~pc, and non-dimensional parameters $\varepsilon_t=0.3$, $\varepsilon_m=0.2$, and $\varepsilon_w=0.01$ (run II) at  $t\approx 1.0\,t_{\rm fmw}$. {\it Panel (a)}: Region with $r\leq 0.8\,R_0$ and $z\leq 0.4\,R_0$ containing the entire cloud. Blue line corresponds to the border of the cloud. {\it Panel (b)}: Region near the primary disk. {\it Panel (c)}: Central part of the cloud. Green line shows the boundary \textcolor{red}{of} the quasi-magnetostatic primary disk and the boundary \textcolor{red}{of} the first hydrostatic core.} 
   \label{fig:rho_hierarchy}
\end{figure}

In Figure~\ref{fig:rho_hierarchy}, we present the structure of the collapsing PSC at the end of the dynamical collapse, $t\sim 1\,t_{\rm fwm}$, when typical picture of the collapse is established. Figure~\ref{fig:rho_hierarchy} shows that hierarchical structure of the cloud is formed. The hierarchy consists of an optically thin geometrically thick flattened envelope, a geometrically and optically thin quasi-magnetostatic, $v_{\rm z} \ll v_{\rm r}$, primary disk inside the envelope, and optically thick first core in the center of the primary disk.
The boundary of the quasi-magnetostatic primary disk is characterized by a sharp jump in the velocity profile $v_z(z)$, when almost free fall of gas along the $z$-direction turns to magnetostatic equilibrium, $v_z\approx 0$. Specifically, we determine the primary disk as a region near the equatorial plane characterized by $v_z<\left(c_{\rm s}^2+v_{\rm A}^2\right)^{1/2}$, where $c_{\rm s}$ is the sound speed, and $v_{\rm A}$ is the Alfven speed. The first core is a hydrostatic region, at the boundary of which both $v_r$ and $v_z$ abruptly decrease below the value of local sound speed~$c_{\rm s}$.

After the formation of the primary disk, a fast MHD shock wave moves from its boundary nearly upwards (see Figure~\ref{fig:rho_hierarchy}(b)). Magnetic field lines are bent and strong toroidal magnetic field is generated behind the shock wave front.  

Soon after the formation of the first core, quasi-magnetostatic equilibrium is violated near its boundary, and an outflow arises in this region, propagating along the lines of the magnetic field (see Figure~\ref{fig:rho_hierarchy}(c)). 

According to our simulations, the magnetic field geometry changes through the internal hierarchy of the cloud. The magnetic field  has a quasi-radial geometry, $B_r\sim B_z$, in the envelope. It remains quasi-uniform, $(B_r,\, B_\varphi)\ll B_z$, inside the primary disk. The magnetic field acquires a quasi-toroidal geometry, $B_\varphi \sim B_z$, in the region between the boundary of the primary disk and the fast MHD shock wave front. As~\cite{Kargaltseva2021} have shown, the specific angular momentum is accumulated near the boundary of the primary disk and then transferred to the envelope by the magnetic braking. The region of efficient magnetic braking lies between the boundary of the primary disk and fast MHD shock wave front. 

The picture of the collapse discussed above demonstrates the leading role of the primary disks in the evolution of the collapsing PSC at the initial stages of collapse.
The primary disk is the main reservoir of mass, angular momentum and magnetic flux feeding the first core and thus determining the characteristics of the protostellar disk forming during further evolution of the system. 

\subsection{Influence of the magnetic field on the dynamics of the collapse}
\label{sec:mf}

Let us analyze the effect of the magnetic field on the dynamics of the collapse of rotating magnetic PSCs, paying special attention to the primary disks. 

\begin{figure*}[htb!]
   \centering
   \includegraphics[width=0.99\textwidth]{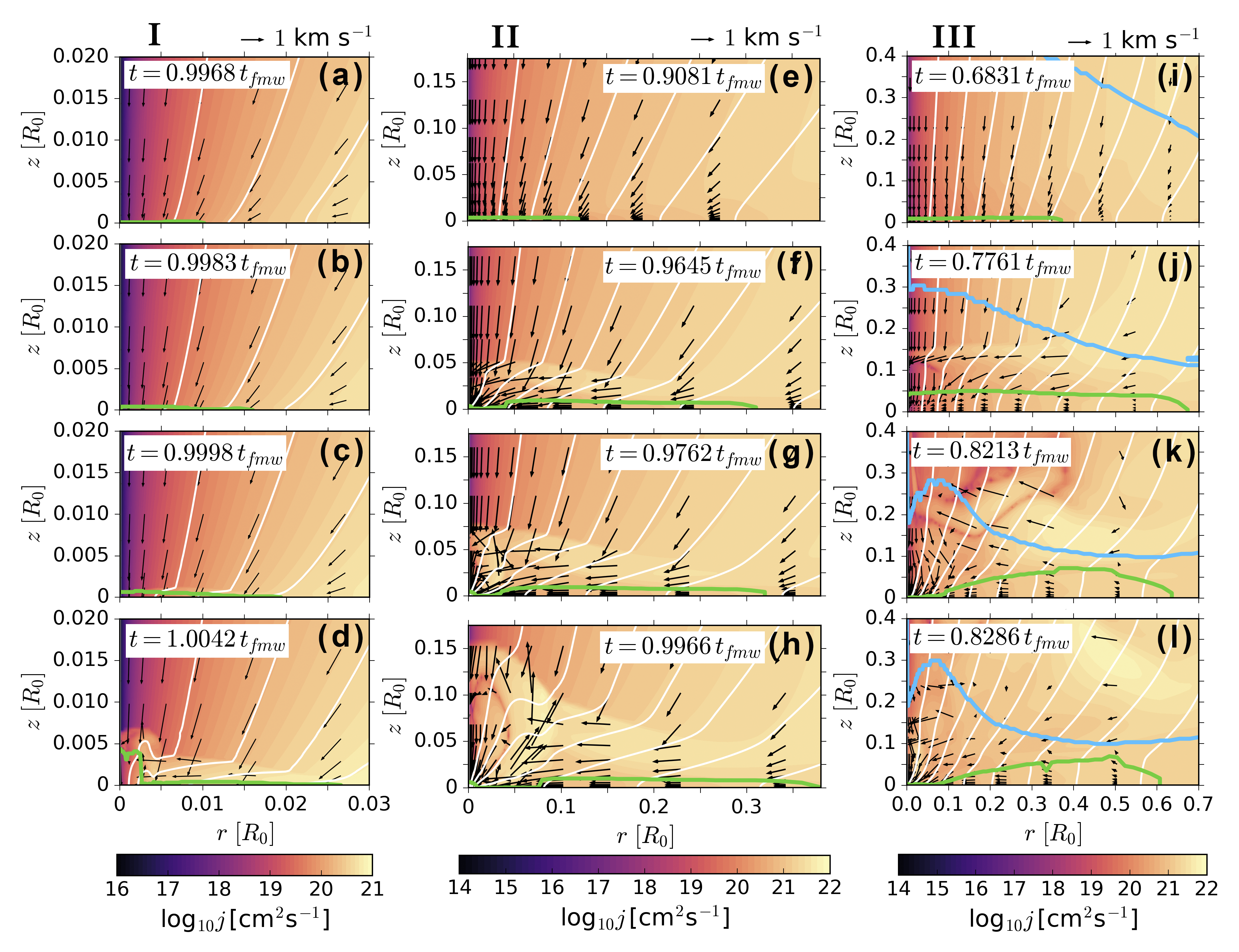}
   \caption{Distribution of the specific angular momentum (color filling), velocity field (arrows) and poloidal magnetic field (white lines) in the simulations with $\varepsilon_m=0.01$ (run I, left panels), $\varepsilon_m=0.2$ (run II, middle panels), $\varepsilon_m=0.6$ (run III, right panels) at different time moments after the primary disk formation (panels from top to bottom). The region near the primary disk is considered. Green line shows the border of the primary disk, blue line is the border of the cloud. } 
   \label{fig:J_PD}
\end{figure*}

In Figure~\ref{fig:J_PD}, we demonstrate structure of the PSC in runs I (left panels), II (middle panels), and III (right panels) at four different time moments after the formation of the primary disk. We plot two-dimensional distribution of the specific angular momentum in order to investigate the efficiency of magnetic braking in each run.

Figure~\ref{fig:J_PD} shows that the hierarchical structure of the collapsing PSC discussed in Section~\ref{sec:hierarchy} forms in all considered runs. During the collapse, primary disks grow in size. For example, the radius of the primary disk in run I increases from $R_{\rm pd}\approx 0.01\, R_0\approx 200$~au to $\approx 0.025\, R_0\approx 500$~au (see Figures~\ref{fig:J_PD}(a, b, c, d)). In all cases the angular momentum firstly accumulates near the surface of the primary disk and then it is transferred to the envelope due to the magnetic braking in the region behind the fast MHD shock wave front. Outflow launching from the region near the first core is also a general feature of considered cases. Typical speed of the outflow is of $2$~km~s$^{-1}$.

The characteristics of primary disks, dynamics of the outflow and efficiency of the magnetic braking changes with $\varepsilon_{\rm m}$. The radius and typical half-thickness of the primary disk increase with $\varepsilon_{\rm m}$. This reflects the increasing role of the electromagnetic force in establishing the magnetostatic equilibrium inside the central part of the cloud. For example, final radius of the primary disk increases from $\approx 0.025\, R_0$ in run I up to $\approx 0.6\, R_0$ in run III (compare bottom panels in Figure~\ref{fig:J_PD}). Therefore, the cloud with strong magnetic field evolves into a state of magnetostatic equilibrium practically as a whole.

The size of the region of efficient magnetic braking bounded by the front of the fast MHD shock wave increases with $\varepsilon_{\rm m}$. Figure~\ref{fig:J_PD}(d) shows that the shock wave travels small distance of $2\times 10^{-3}\,R_0$ along the $z$-direction within the typical dynamical time $t\sim 1\, t_{\rm fmw}$ in run I, while the shock wave in run III travels a distance of $0.3-0.4\,R_0$ and comes out from the envelope by the time $t\sim 0.83\,t_{\rm fmw}$ (see Figure~\ref{fig:J_PD}(l)). This is explained by the fact that the speed of fast MHD shock wave increases with $\varepsilon_{\rm m}$, i.~e. with initial magnetic field strength of the cloud.

The size of the outflow region by the end of the dynamical collapse also increases with increasing role of the magnetic field, since the outflow is driven by the electromagnetic force.

\begin{figure*}[htb!]
   \centering
   \includegraphics[width=0.99\textwidth]{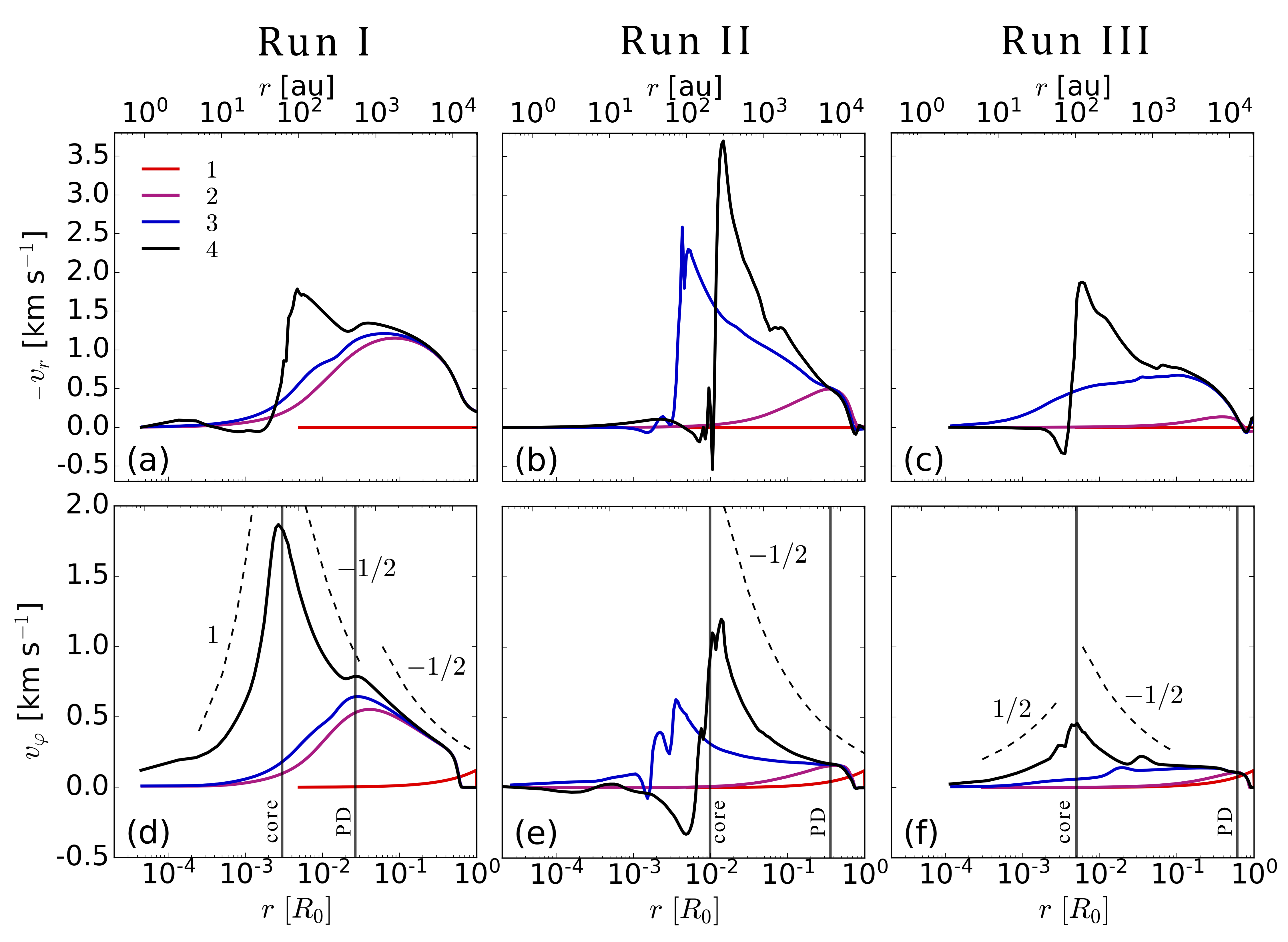}
   \caption{Profiles of radial velocity ($-v_{\rm r}$) and azimuthal velocity $v_{\varphi}$ along the equatorial plane for runs with weak (Run I, the first column), moderate (Run II, the second column), and strong (Run III, the third column) magnetic fields at different moments of time. Line 1 (red color): start of the collapse. Line 2 (pink color): the moment of primary disc formation corresponding to the figures~\ref{fig:J_PD}~(a,~e,~i). Line 3 (blue color): the moment of the first core formation corresponding to the figures~\ref{fig:J_PD}~(c,~f,~k). Line 4 (black color):  the end of the simulation corresponding to the figures~\ref{fig:J_PD}~(d,~h,~l). Dashed lines with numbers show characteristic slopes. Vertical lines show the boundaries of the first core and primary disk at the end of the simulation. } 
   \label{fig:v_r_phi}
\end{figure*}

In order to analyze the dynamics of the collapse in more detail, we plot  the profiles of the radial and azimuthal velocities along the equatorial plane for runs I, II, and III~in Figure ~\ref{fig:v_r_phi}.

At the initial moment of time (lines 1), the cloud rotates rigidly, so that the azimuthal velocity increases with distance as $v_{\varphi} \propto r$. The cloud is at rest in the radial direction initially, $v_{r}=0$. Consider run I with weak magnetic field (Figures~\ref{fig:v_r_phi}a and d). By the time of the formation of the primary disk, differential rotation is established in the region from disk’s boundary, $r\approx 0.02 R_0$, to the periphery of the cloud (line 2). The region of differential rotation increases further in time (line 3). The radial velocity is practically zero inside the first core, $v_{\rm r}\approx 0$, and the entire core rotates rigidly, $v_{\varphi} \propto r$, by the end of the simulation (line 4). The envelope and the primary disk rotate differentially. The azimuthal velocity profile is close to the Keplerian law $v_{\varphi} \propto r^{-1/2}$ in both primary disk and envelope. There is prominent boundary between the primary disk and the envelope seen in $v_\varphi(r)$ profile at $r\approx 0.02\,R_0$. Keplerian rotation profile implies that the first core becomes dominant source of the gravity in the system. 

The azimuthal velocity becomes comparable to the radial one at the boundary of the first core, $v_{\rm r} \approx v_{\varphi}\approx 1.9$~km~s$^{-1}$.

In runs with moderate magnetic field (Figures~\ref{fig:v_r_phi}b and e) and strong magnetic field (Figures~\ref{fig:v_r_phi}d and f), the velocity profiles are different from the velocity profiles in the case of weak magnetic field.

In run II , the first core almost does not rotate by the end of the simulation. The region between the first core and the primary disk, $10^{-3} \lesssim r \lesssim 7\times 10^{-3}\,R_0$, has transient reversed rotation, which is caused by the excitation of the torsional Alfv{\' e}n wave in this region. As in the case of weak magnetic field, the primary disk rotates with Keplerian speed, although the speed is smaller than in Run I. On the contrary, the radial speed is larger with maximum value of $3.5$~km~s$^{-1}$ near the surface of the first core, $r\approx 0.02\,R_0$.

In run~III, the first core rotates differentially, but not in solid-state by the end of the simulation. Inner part of the primary disk rotates with Keplerian speed, while the outer part has almost constant azimuthal velocity. The rotation speed is minimum as compared to Runs I and II, which implies very efficient magnetic braking of the disk.

In both runs II and III $v_{\rm r} > v_{\varphi}$, that is, the centrifugal barrier has not yet formed. Thus, as the magnetic parameter $\varepsilon_{\rm m}$ increases, magnetic braking becomes more efficient.

\subsection{Role of dissipative MHD effects}
\label{sec:diff}

In this section, we analyze the influence of ambipolar diffusion and Ohmic dissipation on the magnetic field strength in the collapsing PSC. We consider run II as a reference one. In order to investigate the role of the dissipative MHD effects in the evolution of the cloud, we performed runs within the ideal MHD limit, as well as taking into account  Ohmic dissipation, magnetic ambipolar diffusion and both types of diffusion.

\begin{figure}
   \centering
   \includegraphics[width=0.5\textwidth]{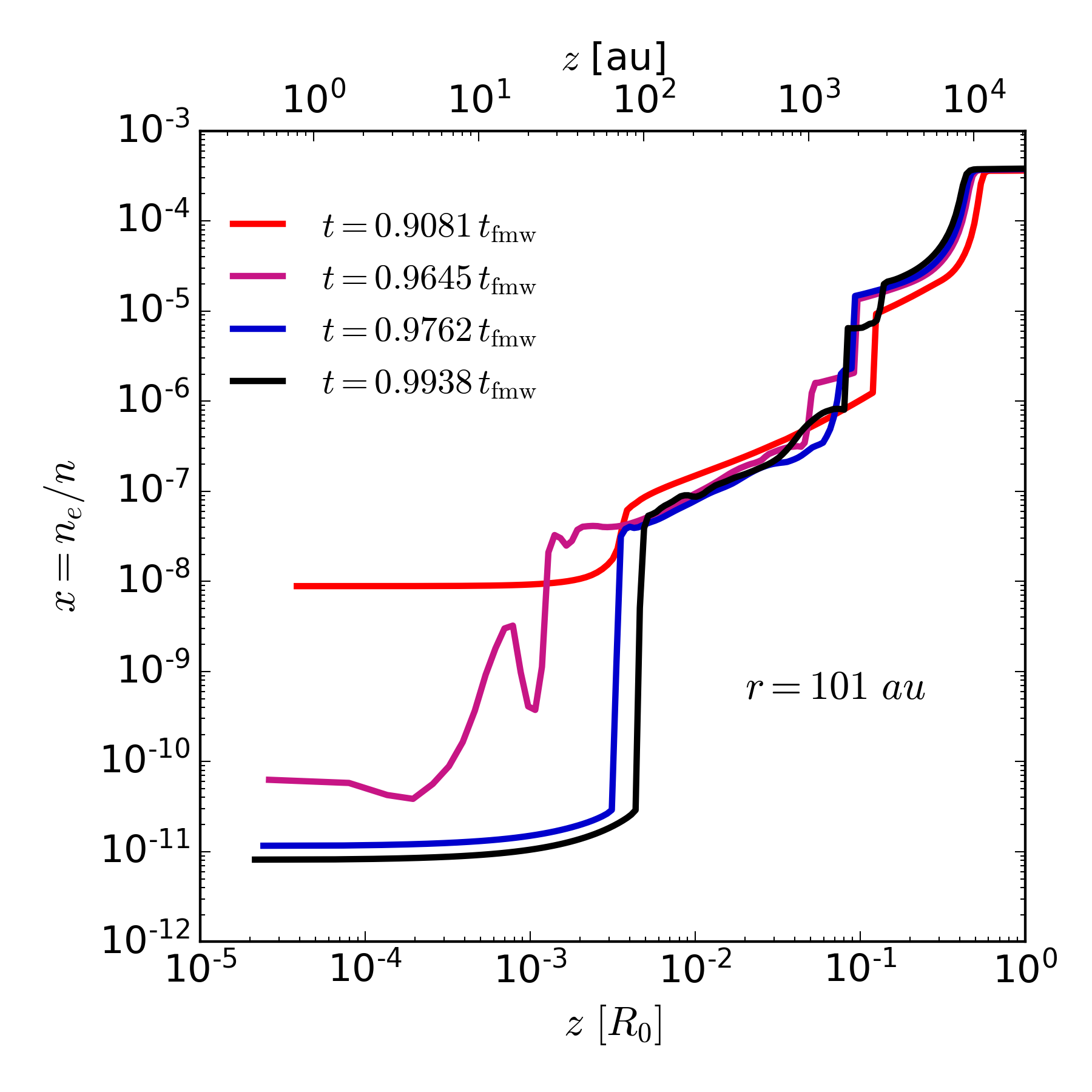}
   \caption{Profiles of the ionization fraction, $x$, along the $z$-direction at $r=100$~au in run II at several times corresponding to the Figures~\ref{fig:J_PD}(e, f, g, h).} 
   \label{fig:X3}
\end{figure}

The efficiency of magnetic diffusion depends on the level of ionization. In Figure~\ref{fig:X3}, we plot the vertical profiles of the ionization fraction, $x$, along the $z$-direction at $r=100$~au for run II at the same times, as in panels (e, f, g, h) of  Figure~\ref{fig:J_PD}. Considered radial distance $r=100$~au corresponds to the region inside the first core close to its boundary. Figure~\ref{fig:X3} shows that the ionization fraction increases with $z$ up to a maximum value of $10^{-5}$ at the cloud periphery. This is due to the decrease in gas density and corresponding increase in the intensity of the interstellar cosmic and X-rays ionization rate. Minimum value of $x$ corresponds to the maximum density in the center of the cloud. Before the formation of the first core at $t=0.9081\,t_{\rm fmw}$ (red line), $x\sim 10^{-8}$ in the central part of the cloud. After the formation of the first core, the degree of ionization drops rapidly down to nearly constant value $x\lesssim 10^{-11}$ inside the core. Therefore, the first core becomes a `dead' zone with small ionization fraction, in which ambipolar diffusion and Ohmic dissipation weaken the magnetic field. The jumps in the profiles of the degree of ionization in the region $z=10^{-3} - 2\times 10^{-2}\,R_0$, correspond to the region of the outflow. Our simulations show that the center of the first core is characterized by smaller ionization fraction of $10^{-13}$.

\begin{figure}
   \centering
   \includegraphics[width=0.5\textwidth]{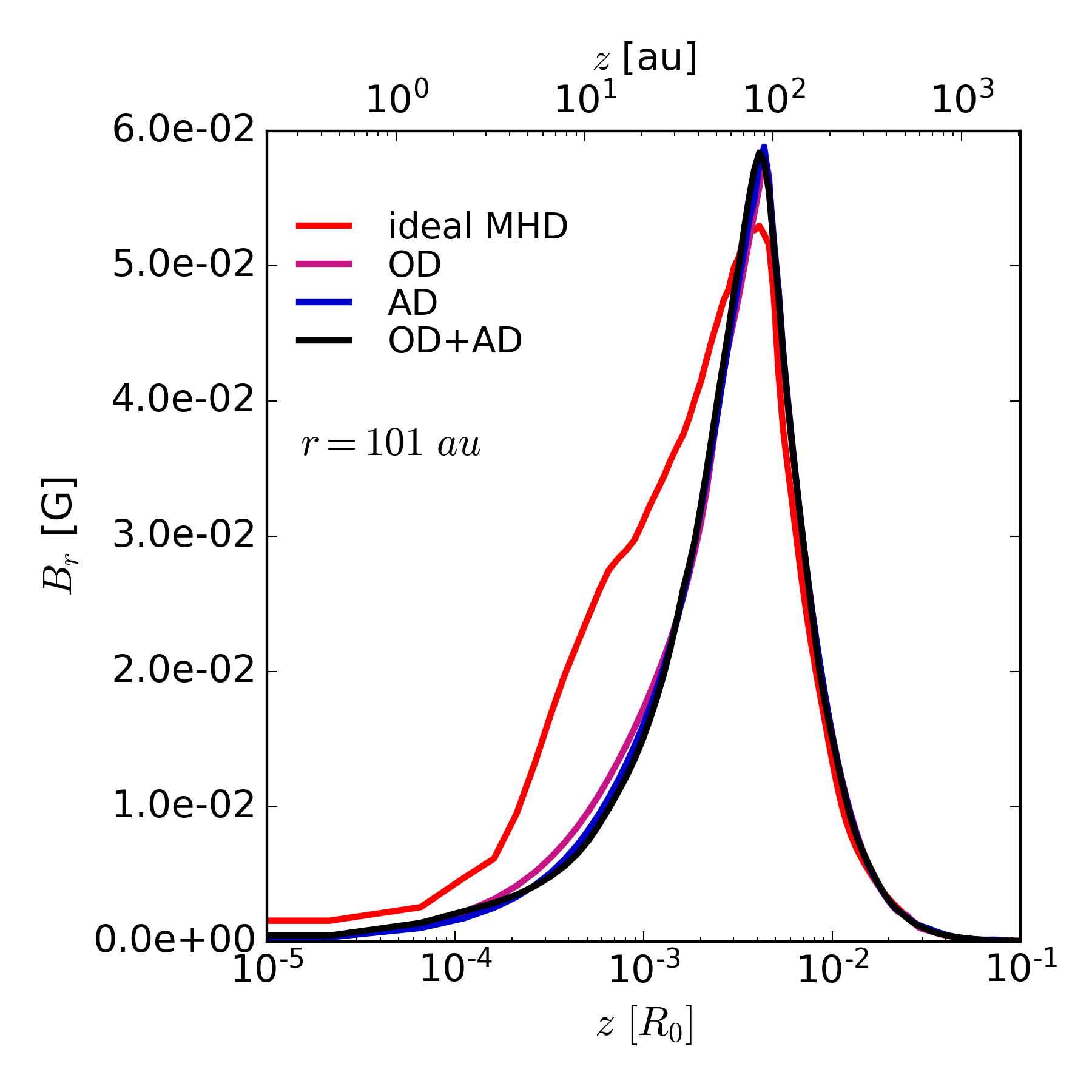}
   \caption{Vertical profiles of the radial component of the magnetic field $B_r$ at 100~au in run with $\varepsilon_t=0.3$, $\varepsilon_m=0.2$ and $\varepsilon_w=0.01$. Different MHD effects are considered: ideal MHD (red line), Ohmic dissipation (pink line), magnetic ambipolar diffusion (blue line) and both types of diffusion (black line).} 
   \label{fig:Br}
\end{figure}

In Figure~\ref{fig:Br}, we plot corresponding vertical profiles of the radial component of the magnetic field $B_r$ at $r=100$~au inside the first core for runs with different MHD effects included. Figure~\ref{fig:Br} shows that, in all four cases, $B_r$ increases from  zero value at the equatorial plane, $z=0$, up to a maximum value of $6 \cdot 10^{-2}$~G at $z\sim 100$~au and then decreases further with $z$. Maximum in the $B_r$ profiles corresponds to the surface of the first hydrostatic core, $z\approx 100$~au.

Inside the first core, the  strength of $B_r$ is larger in the case of ideal MHD comparing to the runs with non-ideal MHD effects. This demonstrates that ambipolar diffusion and Ohmic dissipation weaken the magnetic field in this `dead' zone. This difference will  increase with time as the ionization fraction inside the `dead' zone will drop down during further evolution of the system. Above the first core, the profiles $B_r(z)$ are the same in all runs meaning that the magnetic field is frozen into gas in the envelope.

\section{Conclusions and discussion}

We performed numerical simulations of the collapse of rotating magnetic PSCs with mass of~$10\,M_\odot$ up to the formation of the first hydrostatic core. The simulations were carried out for the cases with weak, moderate and strong magnetic field to analyse influence of the magnetic field on the dynamics of the collapse. The effect of Ohmic dissipation and magnetic ambipolar diffusion on the collapse was analyzed.

The simulations have shown that the formation of the hierarchical structure of the PSC found in our previous works~\citep{Khaibrakhmanov2021, Kargaltseva2021} is the universal property of the collapsing rotating magnetic PSCs. The hierarchy consists of flattened cloud's envelope with the magnet\textcolor{red}{o}static primary disk inside. The first core forms in the central part of the primary disk. The primary disk acts as a reservoir of mass, angular momentum and magnetic flux for the protostar and further forming protostellar disk. Therefore, the characteristics of primary disks are of great importance from the point of view of protostellar disks formation.

Properties and evolution of the structures at each level of the hierarchy strongly depends on the initial magnetic energy of the cloud. The radius of the primary disk increases with $\varepsilon_{\rm m}$ from 500~au at $\varepsilon_{\rm m}=0.01$ to 14000~au at $\varepsilon_{\rm m}=0.6$\textcolor{red}{.} The thickness of the primary disk also increases with $\varepsilon_{\rm m}$, although the primary disk always remain geometrically thin under considered parameters.  The radii of the primary disks are consistent with the observed sizes of flattened envelopes of class 0 YSO~\citep{Ohashi2014, Dunham2014, Tobin2020, Maureira2020}. 

The efficiency of the magnetic braking increases with the initial magnetic energy of the cloud. The region of magnetic braking lies behind the front of the fast MHD shock wave propagating out of the primary disk's surface along the initial magnetic field direction. The magnetic field lines are bent behind the shock front and strong toroidal magnetic field is generated, which drives the magnetic braking. This region grows in time as the shock wave travels into the envelope. In the case of weak initial magnetic field, $\varepsilon_{\rm m}=0.01$, the fast MHD shock wave travels only small distance of $10^{-3}\,R_0$ above the primary disk within the dynamical time scale of the collapse $t_{\rm fmw}$, i.~e. the angular momentum practically does not transferred from the central part of the cloud to its envelope. In the case of moderate initial magnetic field, $\varepsilon_{\rm m}=0.2$, the fast MHD shock wave passes the distance of $0.1\,R_0$ and transfers part of the angular momentum from the primary disk to the envelope within the dynamical time scale. The speed of the fast MHD shock wave is so large in the case of strong magnetic field, $\varepsilon_{\rm m}=0.6$, that its front comes out of the cloud's envelope at $t\sim 0.8\,t_{\rm fmw}$, i.~e. earlier than the end of the dynamical collapse ($t\sim 1\,t_{\rm fmw}$). Significant part of the angular momentum is transported out from the cloud in this case.

Rotational evolution of the cloud depends significantly on the $\varepsilon_{\rm m}$. In the case of weak magnetic field, $\varepsilon_{\rm m}=0.01$ (non-dimensional mass-to-flux ratio $10.5$), following picture establishes after the first core formation. The first core rotates rigidly. Centrifugal barrier appears at its boundary with $v_\varphi\sim v_r\approx 1.9$~km~s$^{-1}$. The primary disk rotates with the Keplerian velocity, as well as the envelope does, which implies that the first core is the dominant source of gravity in the system. In the case of moderate magnetic field, $\varepsilon_{\rm m}=0.2$ (mass-to-flux ratio $2.4$), the first core is practically non-rotating. The region between the first core and the primary disk is characterized by transient reversed rotation because of torsional Alfv{\' e}n wave excitation. The primary disk rotates with Keplerian speed, but slower than in the case of weak magnetic field. In the case of strong magnetic field, $\varepsilon_{\rm m}=0.6$ (mass-to-flux ratio $1.4$), the first core rotates differentially, but it is not in solid-state rotation. Inner part of the primary disk slowly rotates with Keplerian speed, while its periphery rotates with almost constant azimuthal speed. This is a consequence of efficient magnetic braking of the disk's rotation.

According to our simulations, the size of the outflow region at the end of the dynamical collapse increases with $\varepsilon_{\rm m}$. In the weak field case, the outflow region occupies only small volume near the core, while in the strong field case the outflow front propagates practically through the whole cloud by the end of the dynamical collapse. This means that the characteristic time of the outflow development and propagation from the cloud's center decreases with $\varepsilon_{\rm m}$. This is due to the increasing role of the electromagnetic force driving the outflow.

The `dead' zone with the low ionization fraction, $x\leq 10^{-13}-10^{-11}$, forms inside the first core and at the base of the outflow region. In this region, Ohmic dissipation and magnetic ambipolar diffusion weaken the magnetic field comparing to the ideal MHD case, This effect is most pronounced for the radial component of the magnetic field.
This result is consistent with early studies of the influence of non-ideal MHD effects at the initial stages of collapse~\citep[e.g., ][]{BlackScott1982, DudSaz1987, Mouschovias1991}. Evolution of the first core and primary disk, as well as properties of the further forming protostellar disks in clouds with different initial magnetic energies, will depend on the efficiency of the Ohmic dissipation and magnetic ambipolar diffusion in this region. 

In future we plan to develop our approach and investigate the hierarchy of the collapsing rotating magnetic PSCs in application to more realistic initial configurations of the cloud. Construction of the synthetic continuum emission maps and polarization maps of collapsing PSCs on the basis of our simulations will allow to interpret observations of class 0 YSOs and analyze the conditions for the formation of protostellar disks.

\subsection*{Acknowledgements}
This work is financially supported by the Russian Science Foundation (project 19-72-10012). The authors thank anonymous referee for useful comments.

\end{document}